\documentclass[useAMS,usenatbib]{mn2e}
\usepackage{amssymb,amsmath,epsfig,times, natbib}
\title[Cavity in A1795]{Exploring the origin of a large cavity in Abell 1795 using deep \emph{Chandra} observations}
\author[S. A. Walker et al.]{S. A. Walker,$^1$\thanks{Email: 
    swalker@ast.cam.ac.uk} A. C. Fabian$^1$ and P. Kosec$^1$\\
  $^1$Institute of Astronomy, Madingley Road, Cambridge CB3 0HA \\
  \\
    \\
   \\
   \\
}
\date{}

\begin{document}

\maketitle

\begin{abstract}
We examine deep stacked \emph{Chandra} observations of the galaxy cluster Abell 1795 (over 700ks) to study in depth a large (34 kpc radius) cavity in the X-ray emission. Curiously, despite the large energy required to form this cavity ($4PV=4\times10^{60}$ erg), there is no obvious counterpart to the cavity on the opposite side of the cluster, which would be expected if it has formed due to jets from the central AGN inflating bubbles. There is also no radio emission associated with the cavity, and no metal enhancement or filaments between it and the BCG, which are normally found for bubbles inflated by AGN which have risen from the core. One possibility is that this is an old ghost cavity, and that gas sloshing has dominated the distribution of metals around the core. Projection effects, particularly the long X-ray bright filament to the south east, may prevent us from seeing the companion bubble on the opposite side of the cluster core. We calculate that such a companion bubble would easily have been able to uplift the gas in the southern filament from the core. Interestingly, it has recently been found that inside the cavity is a highly variable X-ray point source coincident with a small dwarf galaxy. Given the remarkable spatial correlation of this point source and the X-ray cavity, we explore the possibility that an outburst from this dwarf galaxy in the past could have led to the formation of the cavity, but find this to be an unlikely scenario.  
 
\end{abstract}


\section{Introduction}

Cavities in the intracluster medium (ICM) have been extensively studied by \emph{Chandra}, whose unrivalled subarcsecond spatial resolution has allowed the depressions in X-ray emission caused by the cavities to be revealed for large samples of clusters for the first time. These cavities are blown into the ICM on either side of the brightest cluster galaxy (BCG) by jets from the central active galactic nucleus (AGN), and provide a unique and invaluable way of measuring the amount of energy dissipated into the ICM from AGN feedback. These detailed observations have revealed that in the cores of cool core clusters, where the cooling time is much shorter than the Hubble time, the cooling of X-ray emitting gas is on average balanced by feedback from the central AGN, preventing the rapid cooling of gas onto the central BCG and limiting the rate of star formation.

In the cool core cluster Abell 1795, the cavity structure is unusual. A large depression to the north west was identified in \citet{Crawford2005}, which curiously has no counterpart on the other side of the cluster. The cavity has no radio emission inside it, which is typically seen for bubbles inflated by jets from central AGN (e.g. Perseus, \citealt{Fabian2006}; MS 0735.6+7421, \citealt{McNamara2009}) due to synchrotron radiation from the relativistic electrons filling them. This cavity also does not appear to have dragged behind it any H$\alpha$ filaments if it has risen from the central AGN, in sharp contrast to what has been observed in other clusters (e.g. Perseus). The prominent, X-ray bright H$\alpha$ filament to the south of the cluster core terminates at a cavity which is unusually extremely small given its distance from the central AGN. 

Recently, \citet{Maksym2013} and \citet{Donato2014} have both independently identified a time variable X-ray point source lying inside the northwestern cavity, associated with the dwarf galaxy WINGS J134849.88+263557.5 (hereafter WINGS J1348). This source is very luminous in early \emph{Chandra} observations taken in 1999, but has since faded over the last decade and is not visible in the \emph{Chandra} data taken since 2004. The source cannot be identified in any earlier observations taken with previous X-ray observatories. \citet{Maksym2013} and \citet{Donato2014} conclude that this flare in X-ray luminosity is due to a tidal disruption event. The coincidence of this source with the X-ray cavity (it lies very near to the centre of the cavity), is very curious. 

Subsequent follow up Gemini spectroscopy by \citet{Maksym2014} has confirmed that the redshift of WINGS J1348 is the same as that of Abell 1795, and that the galaxy is indeed in the cluster as opposed to being a foreground or background galaxy. \citet{Maksym2014} conclude that WINGS J1348 is an extremely low mass dwarf galaxy ($M_{\rm gal}=3\times10^{8}$M$_{\odot}$), and limit the black hole mass to log$[M_{\rm BH}/{\rm M_{\odot}}] \sim 5.3- 5.7$.

Here we stack and analyse the catalogue of \emph{Chandra} observations taken of Abell 1795 to understand the dynamics of the northwestern cavity and its origin. We also discuss the viability of possible scenarios which could connect the cavity with the variable X-ray point source located inside it.

%T profile across cavity edge.
%T profile across cold front in sectors. 
%Energy in cavity.
%Metal distribution.
%

\begin{figure*}
  \begin{center}
    \leavevmode
    \hbox{
      \epsfig{figure=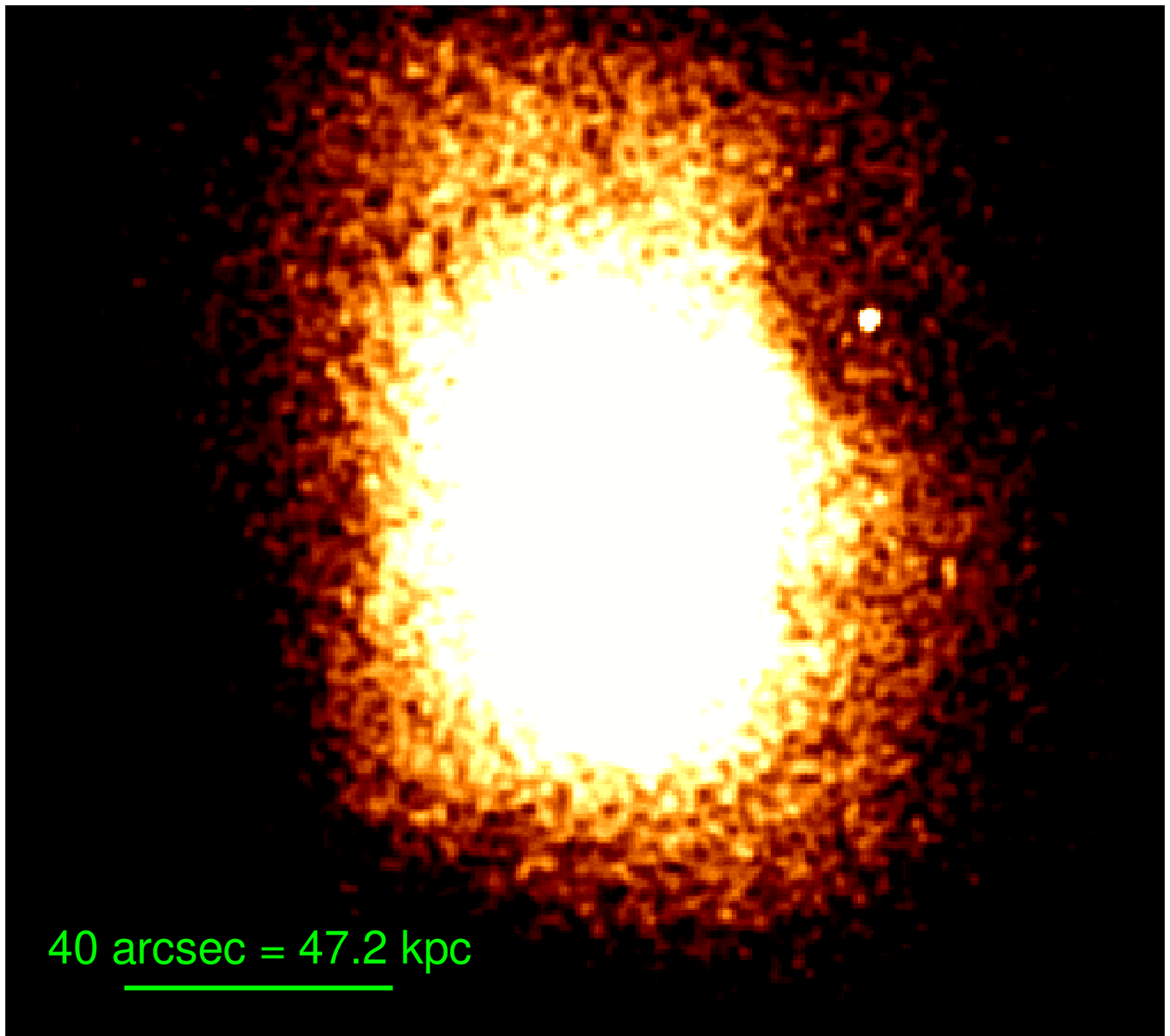,
        width=0.45\linewidth}
              \epsfig{figure=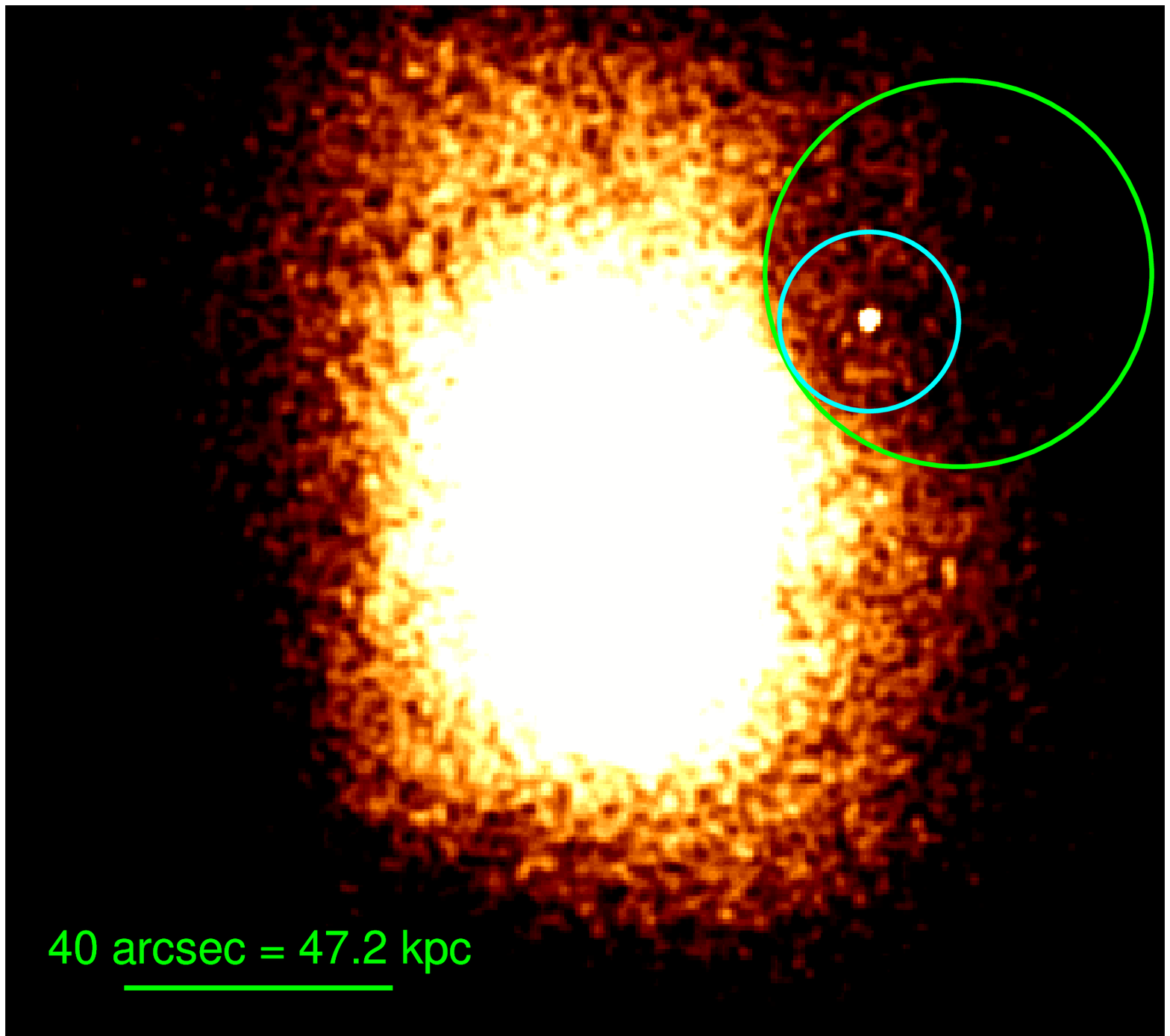,
        width=0.45\linewidth}
            }
       \hbox{
      \epsfig{figure=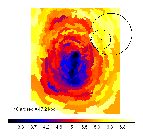,
        width=0.45\linewidth}
            }
      \caption{\emph{Top left}: Smoothed image of stacked ACIS-S and ACIS-I data for Abell 1795, showing the inner rim of the cavity to the north west \emph{Top right}: Same as the top left hand panel but showing circles centred on the point source (cyan circle) and the centre of curvature of the cavity (green circle), which have radii of 15.8 kpc and 34 kpc respectively.   \emph{Bottom left}: Temperature map using regions with a signal to noise of 100, showing that the cavity has a higher temperature than the surrounding gas at the edge closest to the core. }
      \label{image}
  \end{center}
\end{figure*}

 We use a standard $\Lambda$CDM cosmology with $H_{0}=70$  km s$^{-1}$
Mpc$^{-1}$, $\Omega_{M}=0.3$, $\Omega_{\Lambda}$=0.7. For the redshift of Abell 1795, $z=0.0625$, this yields an angular scale of 72.2 kpc/arcmin. All errors unless
otherwise stated are at the 1 $\sigma$ level. 

\begin{figure*}
  \begin{center}
    \leavevmode
    \hbox{
      \epsfig{figure=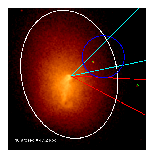,
        width=0.43\linewidth}
              \epsfig{figure=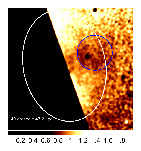,
        width=0.42\linewidth}
        }
         \hbox{
         \epsfig{figure=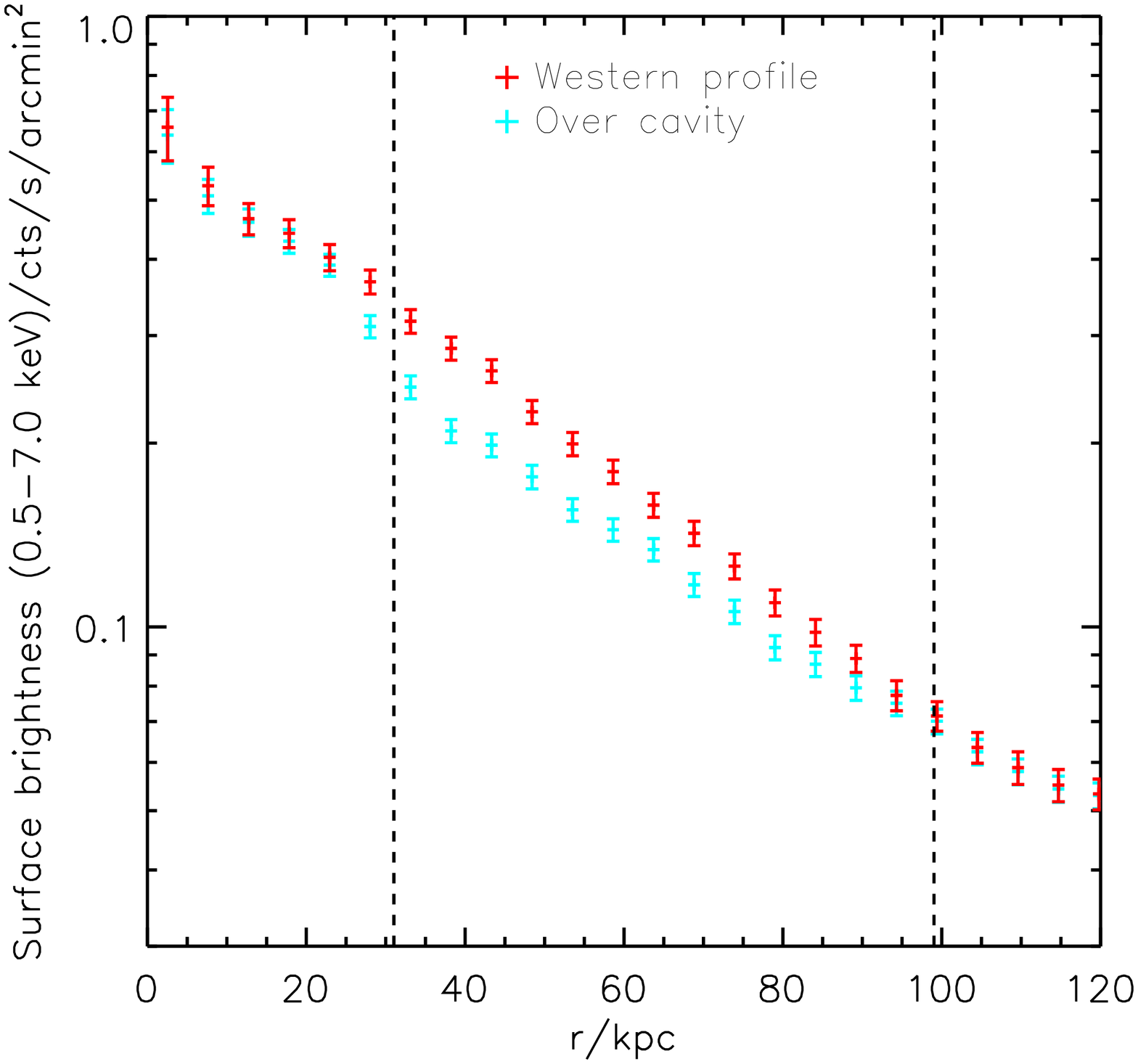,
        width=0.45\linewidth}      
                    }
  
      \caption{\emph{Top Left}: Image of the cluster showing the elliptical nature of the overall cluster emission, with the white ellipse showing the approximate shape of an isophote. We find the azimuthal residuals using elliptical annuli concentric with the one shown in the left panel, and these residuals are shown in the \emph{top right} hand panel, showing the full extent of the north western cavity (denoted by the blue circle in both panels). These residuals were found for the western half of the cluster, so as to exclude the filament to the south east of the core which would produce spurious features in the residuals if included. The \emph{bottom} panel shows surface brightness profiles over elliptical annuli across the cavity and to the south of the cavity, in the directions of the cyan and red sectors respectively shown in the top left hand panel. The vertical dashed lines show the inner and outer edges of the blue, 34 kpc radius circle.     }
      \label{azav_residuals}
  \end{center}
\end{figure*}

\begin{figure*}
  \begin{center}
    \leavevmode
    \hbox{
      \epsfig{figure=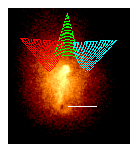,
        width=0.4\linewidth}
              \epsfig{figure=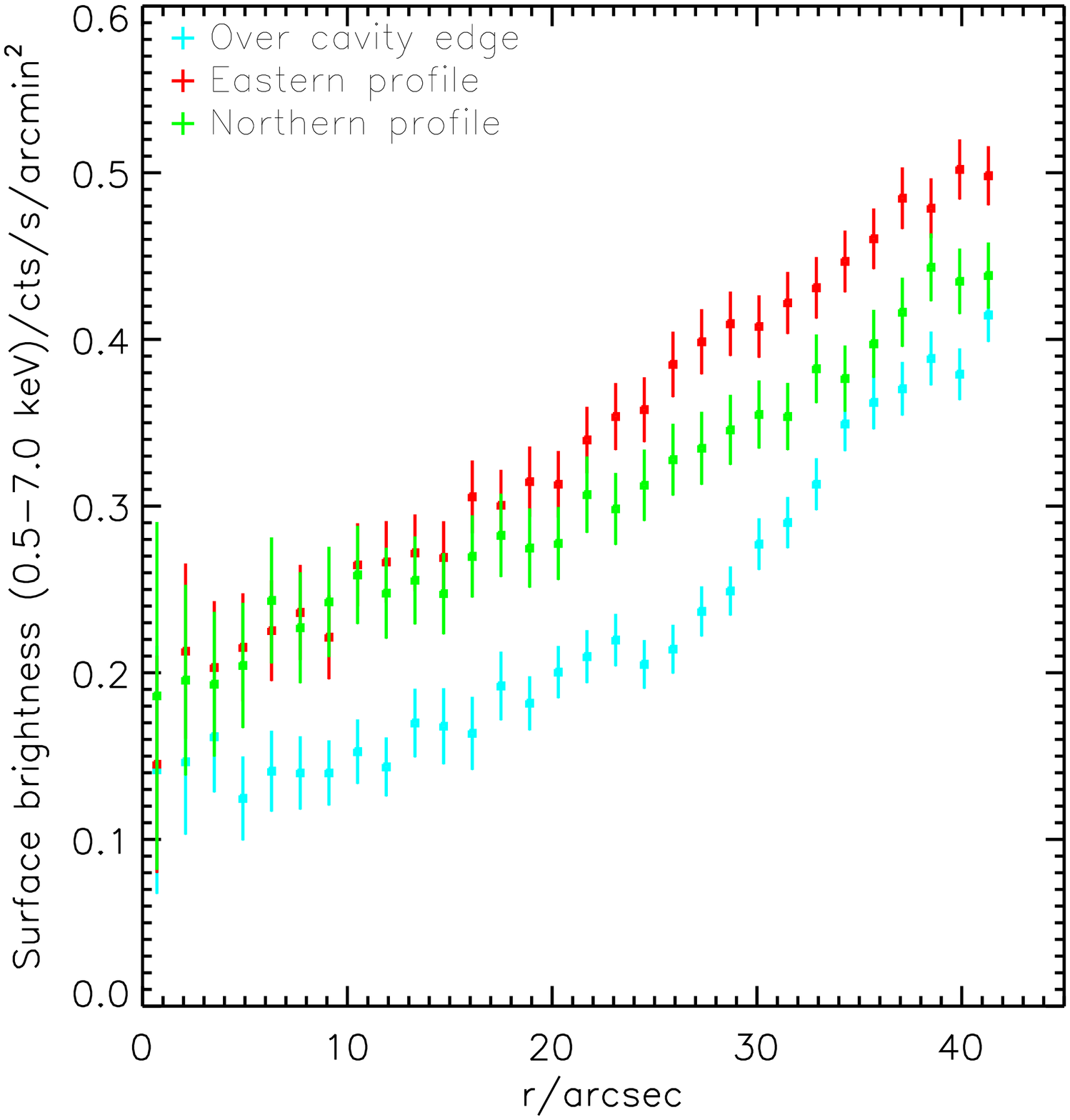,
        width=0.45\linewidth}
            }
  
      \caption{\emph{Left}: Shows three sectors along which surface brightness profiles were obtained (shown in the right hand panel) to show the drop in surface brightness in the direction of the cavity. All of these profiles start at the same distance from the BCG and move inwards. The annuli were chosen to have the same curvature as the cavity edge. The small southern depression in X-ray brightness identified in \citet{Crawford2005} is shown with the white arrow.}
      \label{sectors}
  \end{center}
\end{figure*}

\section{Observations}

As Abell 1795 has been extensively used as a calibration source, a large archive of \emph{Chandra} observations has accumulated. These data have been obtained for a variety of aim points, with some having the core at the centre of the field of view, and others having the core near the edge of the field of view. As we require high spatial resolution for our analysis of the cavities, we use only the data which has the cluster core on axis, as combining off axis observations will lower the overall spatial resolution due to the increase in the \emph{Chandra} PSF off-axis. The ACIS-S and ACIS-I data used are tabulated in tables \ref{ACIS-S_obs} and \ref{ACIS-I_obs} respectively in appendix \ref{sec:appendix}. The total observation time is 710ks. The off axis data not used are 80 ks of ACIS-I data. Due to the smaller effective area of ACIS-I compared to ACIS-S, and the significant vignetting of these off-axis observations, including them in the later spectral map analysis is found to have a negligible effect.

\section{Data Reduction}

\subsection{Images}

Each events file was reprocessed using \textsc{chandra\_repro} using \textsc{CIAO} 4.6. To create the stacked 0.7-7.0keV image shown in Fig. \ref{image} we used the \textsc{CIAO} script \textsc{reproject\_obs} to reproject the events files, followed by the \textsc{CIAO} script \textsc{flux\_obs} to extract images from the events files in the 0.7-7.0 keV band and make their exposure maps.  We then combined all of the images weighted by their exposure maps. This yielded the stacked, exposure corrected image combining the ACIS-I and ACIS-S data in the 0.7-7.0 keV which is shown in the top two panels in Fig. \ref{image} and the top left hand panel of Fig. \ref{azav_residuals}. In all of the images north is up and west is to the right.

We used the \textsc{CIAO} script \textsc{acis\_bkgrnd\_lookup} to identify and download the suitable blank sky background fields for all of the Abell 1795 events files, which were reprojected to match the coordinate system of the Abell 1795 events files. The background files were scaled so that the count rates in the hard 10-12 keV band (where the effective area is zero and the signal is due to the particle background) matched those of the cluster observations.

\subsection{Spectral analysis}

We added together the ACIS-I and ACIS-S counts images and subtracted the appropriately scaled mosaicked blank sky background field to obtain a background subtracted counts image. Point sources were identified using \textsc{wavdetect} and masked from all further analysis. We then used the contour binning method of \citet{Contbin2006} to bin the image into regions with a signal to noise of 100 which follow the X-ray surface brightness contours. Spectra for each region were then extracted from all of the events files using \textsc{dmextract}, with ARFs and RMFs created for each region and for each event file using \textsc{mkwarf} and \textsc{mkacisrmf} respectively. The spectra for each specific region from all of the events file were then combined and weighted appropriately using the ARFs to produce the total spectrum for each region. The spectral extraction was done separately for the ACIS-I and ACIS-S data, so each region had a total ACIS-I spectrum and a total ACIS-S spectrum which were fitted simultaneously in XSPEC rather than added due to differences in the effective areas of the two different detectors. For each region a background spectrum from the blank field backgrounds was also extracted. 

To produce the temperature map shown in the bottom panel of Fig. \ref{image}, each region was fitted with a single temperature absorbed apec component with the temperature, metallicity and normalisation allowed to be free parameters. The column density was fixed to the LAB value from \citet{LABsurvey} of $1.2\times10^{20}$ cm$^{-2}$.

\section{Cavity properties}

\subsection{Image Analysis}

The central rim of the cavity feature to the northwest can be clearly seen in the top left panel of Fig. \ref{image}. The curvature of the rim is consistent with a spherical shaped cavity of radius 34 kpc, shown as the green circle in the top right panel of Fig. \ref{image}. The bright point source located inside the cavity is from the tidal disruption event associated with the dwarf galaxy WINGS J1348 discussed earlier (\citealt{Maksym2013}, \citealt{Donato2014}, \citealt{Maksym2014}), and is only visible in the observations taken before 2004. The galaxy lies 16kpc from the estimated centre of the cavity (assuming it is circular), almost exactly along the direction connecting the BCG and the cavity centre.

The temperature map shown in the bottom panel of Fig. \ref{image} shows the temperature to be much higher inside the inner rim of the cavity (6.3$^{+0.1}_{-0.1}$ keV), where the surface brightness drops abruptly, than outside (5.2$^{+0.1}_{-0.1}$ keV). The curvature of the rim is obviously the wrong way around for this surface brightness discontinuity to be a cold front, and is unlikely to be the result of the sloshing activity which has been explored in \citet{Markevitch2001} and \citet{Ehlert2014}. 

Due to the high surface brightness near the core, the inner rim of the cavity is the easiest to observe, and is clearly visible in the raw images. To explore the full extent of the cavity, instead of just its inner rim, we divide the image by the azimuthal average from elliptical annuli concentric with the ellipticity of the ICM emission, as shown in the top right panel of Fig. \ref{azav_residuals}. Here we only divide by the average from the half of the cluster to the west of the major axis, because the prominent H$\alpha$ filament to the south east makes the eastern side unrepresentative of the overall shape of the cluster emission as a whole. The result is shown in the top right hand panel of Fig. \ref{azav_residuals}, where we see a depression in surface brightness consistent with the volume covered by the spherical cavity we determined earlier from the shape of the inner rim (shown as the blue circle). Due to the rapid decline in X-ray surface brightness with radius, it is challenging to accurately determine the full shape of the cavity, but it appears consistent with the simple assumption of being spherical.  

To explore the cavity extent further, we compare surface brightness profiles along sectors directed radially outwards across the cavity (the cyan region in the top left panel of Fig. \ref{azav_residuals}) and immediately to the south of the cavity (the red region). These profiles are shown in the bottom panel of Fig. \ref{azav_residuals}, showing that the decrement extends out to a radius of 90-100 kpc, roughly in agreement with the assumption that the cavity is a sphere of radius 34 kpc. The vertical dashed lines show the locations of the inner and outer rims of such a sphere for comparison.  

In Fig. \ref{sectors} we compare surface brightness profiles across the cavity rim with similar profiles moving radially inwards from the north and east, which clearly shows the surface brightness discontinuity across the inner cavity rim (the cyan profile).

\subsection{Calculation of cavity energy}

If the cavity has expanded slowly to a volume, $V$, against the pressure, $P$, of the surrounding ICM, the total energy required for the formation of the cavity is the sum of the internal energy within it and the work done in growing the bubble, given by:
\begin{eqnarray}
E = \frac{1}{\gamma_{\rm cav} - 1} PV + PV = \frac{\gamma_{\rm cav}}{\gamma_{\rm cav} - 1}PV
\end{eqnarray}
where mean adiabatic index of the gas inside the cavity, $\gamma_{\rm  cav}$, is $4/3$ for a relativistic plasma, yielding, $E=4PV$. Assuming the cavity to be a sphere of radius 34 kpc, (as shown by the green circle in the top left panel of Fig. \ref{image}), this yields a total energy of $E=4PV=3.8^{+2}_{-1} \times 10^{60}$ erg, where the errors include the systematic error in measuring the pressure due to the pressure gradient across the cavity. The energy contained in the cavity is then $PV=10^{60}$ erg.

We can estimate the age of the cavity using three characteristic timescales: the sound crossing time, $t_{cs}$, the buoyancy rise time $t_{buoyancy}$ (the time taken for the bubble to reach terminal buoyancy velocity), and the refill time, $t_{refill}$, (the time taken for the bubble, starting at rest, to rise buoyantly through its own diameter). Firstly we calculate the sound crossing time, given the distance between the centre of the bubble and the central AGN, $R=54$ kpc, and the sound speed, $c_{s}$:
\begin{eqnarray}
t_{\rm cs} = \frac{R}{c_{\rm s}}
\end{eqnarray}
where
\begin{eqnarray}
c_{\rm s} = \sqrt{\gamma_{\rm ICM} kT / \mu m_{\rm p}}
\end{eqnarray}
where $\gamma_{ICM}=5/3$ is the adiabatic index of the ICM, $\mu$=0.62 is the mean atomic weight and the proton mass is $m_{p}$. This yields a sound crossing time of $t_{cs}=40.9$Myr.

The buoyancy time is given by:
\begin{eqnarray}
t_{\rm buoyancy}= R \sqrt{\frac{S C_{\rm D}}{2gV}}
\end{eqnarray}
where $S=\pi r^{2}$ is the cross-sectional area of the bubble (which has radius $r=34$kpc), the local gravitational acceleration is $g=GM(<R)/R^{2}$, the drag coefficient is $C_{\rm D}=0.75$ (as found in the simulations of \citealt{Churazov2001}), and $V$ is the bubble volume. The buoyancy time is 39.5Myr.

The refill time, given by
\begin{eqnarray}
t_{\rm refill}= 2 \sqrt{\frac{r}{g}}
\end{eqnarray}
is 93.6 Myr. 

We take the age of the cavity to the be the mean of these three timescales (as in \citealt{Vantyghem2014}), which is $t_{\rm age}=$58Myr. 

 The power involved in inflating the bubble can then be estimated as 
\begin{eqnarray}
 P = \frac{E}{t_{\rm age}}
 \end{eqnarray} 
which yields $P=2.1\times10^{45}$ erg s$^{-1}$. This power exceeds by a factor of four the X-ray bolometric luminosity within the cooling radius (which is $L_{\rm X}=5.0\times 10^{44}$ erg s$^{-1}$). Here we define the cooling region as the region within which the cooling time is less than the lookback time to $z=1$ in a standard $\Lambda$CDM cosmology, which is 7.7 Gyr. For Abell 1795 the radius of the cooling region is 100 kpc. Assuming that there is another cavity on the opposite side of the cluster which we cannot see (possibly due to projection effects, as we discuss later), the total power output from the central AGN responsible for the cavity should be roughly twice this, $P_{tot}=4.2\times10^{45}$ erg s$^{-1}$, which would be at least 8 times the X-ray bolometric luminosity within the cooling radius. The vast majority of the cavity volume lies within the cooling region (i.e. within the central 100kpc) allowing this energy to be deposited within the cooling radius. We see that this outburst can in principle comfortably compensate for radiative cooling when the cavity dissipates and releases its enthalpy to heat the ICM.

\subsection{Cavity size and location}
\label{sec:cavitysize}
As found in \citet{Rafferty2006}, the distribution for the ratio between the cavity distance from the BCG, R, and the cavity radius, r, peaks at 2 and has a range between 1 and 3.5. For the north western bubble, the radius is $r$=34 kpc and the nuclear distance is $R$=54kpc, yielding, $R/r$=1.6, which is in good agreement with the observed range of values, adding support to the idea that this cavity has risen from the central AGN.  

For the small southern bubble identified in \citet{Crawford2005} and shown with the white arrow in Fig. \ref{sectors}, which has a radius, $r$, of 3 arcsec (3.5 kpc) and lies at a distance, $R$, of 38 arcsec (45 kpc) from the core, the ratio $R/r$ is abnormally high (it is 12.7).  This feature, if it is a bubble which has risen from the core, represents a significant outlier from the distribution of cluster size and nuclear distance observed in other clusters.

\subsection{Metal distribution}

\begin{figure}
  \begin{center}
    \leavevmode
    \hbox{
      \epsfig{figure=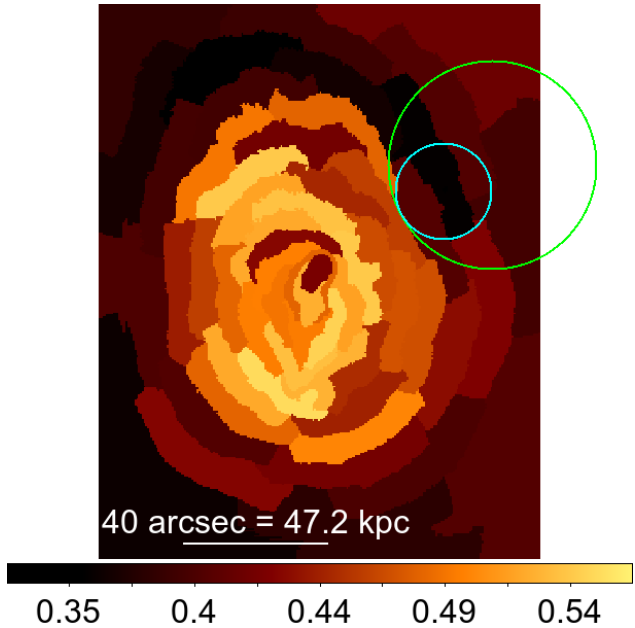,
        width=\linewidth}
        }
          
      \caption{Map of the metal abundance with the cavity location indicated by the circles as in Fig. \ref{image}. The image was binned into regions with a signal to noise of 200, and each region was fit with a single temperature model}
      \label{metals}
  \end{center}
\end{figure}

To explore the metal distribution, we rebinned the image into regions with higher signal to noise (200) using the contour binning method of \citet{Contbin2006}. We fitted the extracted spectra in each region with an absorbed single temperature \textsc{apec} component, with the temperature, abundance and normalisation allowed to be free parameters. The resulting metal abundance map is shown in Fig. \ref{metals}. 

If the northwestern cavity is a bubble which has risen from the centre, we would expect that it could have transported metals from the BCG, leading to an enhancement in metal abundance. Much work (for example \citealt{Simionescu2008}, \citealt{Simionescu2009}, \citealt{Kirkpatrick2009}, \citealt{OSullivan2011} and \citealt{Kirkpatrick2011}) has found that the metal abundance in the cores of galaxy clusters is anisotropic and aligned with large scale radio and cavity axes, indicating that metals are driven into the ICM along outflows driven by the jets from the AGN. 

However, we see in Fig. \ref{metals} that the inner edge of the north western cavity corresponds to a drop in metal abundance, which would appear to be in tension with the idea that the cavity is bubble which has risen from the BCG.

\subsection{Comparing X-ray and Radio data}

\begin{figure*}
  \begin{center}
    \leavevmode
    \hbox{
      \epsfig{figure=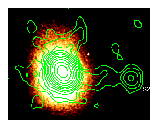,
        width=0.5\linewidth}
        }
        
          \hbox{
              \epsfig{figure=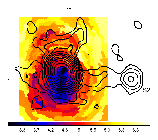,
        width=0.5\linewidth}
                      \epsfig{figure=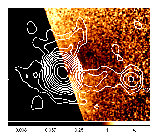,
        width=0.5\linewidth}
            }
  
      \caption{Plotting the 1.4GHz VLA radio contours for Abell 1795 from \citet{Giacintucci2014} over the exposure corrected \emph{Chandra} image (top), the temperature map (bottom left) and the azimuthal residual map (bottom right), showing the correlation between the radio contours and the edge of the north western cavity feature in the X-ray data. The source to the extreme right labelled S2 is a separate radio source unrelated to the minihalo. The 1.4GHz radio contours are 1, 2, 4, 8, 16 $\times$3$\sigma$, and were obtained using the VLA C-array with a resolution of 18.7$''$ $\times$ 16.0$''$ in position angle 37$^{\circ}$.    }
      \label{radiooverplot}
  \end{center}
\end{figure*}

\citet{Giacintucci2014} examined 1.4 GHz VLA radio data for Abell 1795, and found a candidate minihalo. Such a minihalo could be the result of sloshing activity generating turbulence which reaccelerates radio emitting electrons. The VLA 1.4 GHz radio data for Abell 1795 presented in figure 7 of \citet{Giacintucci2014} show that the radio contours appear to be curved around the inner rim of the north western X-ray cavity. In Fig. \ref{radiooverplot} we overplot these radio contours over the exposure corrected \emph{Chandra} image, the temperature map, and the residual map from Fig. \ref{azav_residuals}, showing the remarkable correspondence between the X-ray cavity and the radio contours. There are two extended, possibly filamentary features reaching to the north and west, which appear to `wrap around' the volume of the X-ray cavity. The radio contours appear to be confined behind the surface brightness edge. 

\subsection{Resemblance to features in other clusters}

The inner edge of the northern cavity also bears a resemblance to the southern `bay' feature in Perseus described in \citet{Fabian2011}, which also has a sharp concave surface brightness discontinuity inside of which there is a sudden temperature increase and a `tongue' of hotter material (see the bottom panels of figure 12 from \citealt{Fabian2011}). The southern bay is around twice as far from the cluster centre than the northwestern feature in A1795 (100kpc compared to 54kpc), and has no coincident radio emission. It is possible that the southern bay in Perseus is the result of an old, elliptical ghost bubble (see figure 9 in \citealt{Fabian2011}) whose full extent is difficult to trace due to the rapid decline in X-ray surface brightness with distance from the cluster core. 

\begin{figure*}
  \begin{center}
    \leavevmode
    \hbox{
      \epsfig{figure=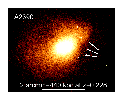,
        width=0.33\linewidth}
         \epsfig{figure=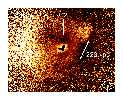,
        width=0.33\linewidth}     
         \epsfig{figure=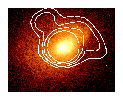,
        width=0.33\linewidth}     
        }  
      \caption{ \emph{Left}: Exposure corrected \emph{Chandra} image of A2390 in the 0.5-7.0 keV band, showing the inner edge of a cavity feature to the west, qualitatively similar to the feature in A1795. \emph{Centre}:Residuals of the image from the azimuthal average using elliptical annuli, showing the depression to the west whose diameter (220kpc) is much larger than that of the feature in A1795. A possible companion cavity, labelled C2, is also visible in this residual image. \emph{Right}:Overplotting the NVSS radio contours over the \emph{Chandra} image, showing that the radio emission extends into the western cavity. 
   }
      \label{A2390}
  \end{center}
\end{figure*}

A similar inner edge feature is also present in Abell 2390 (and is commented on in \citealt{Vikhlinin2006}), but is around 5 times larger than the feature in A1795, as shown in Fig. \ref{A2390}. It is also much further out in the cluster, with the inner rim being 200 kpc from the BCG compared to 54 kpc in A1795. The edge is clearly visible in the exposure corrected \emph{Chandra} image in the 0.5-7.0 keV presented in the left panel of Fig. \ref{A2390}, and its full extent becomes clear in the residual image from an elliptical model shown in the centre panel. A second cavity is also visible, labelled C2. 

However the correspondence between the radio and X-ray data is different for A2390. As shown in the right panel of Fig. \ref{A2390}, the NVSS radio contours extend into the western cavity, and are not constrained by the inner edge of the cavity as in A1795.  

The \emph{Chandra} images of A2390 shown were obtained by merging the available observations, (obsids: 500, 501, 4193), which combined have total exposure of 113ks. The \emph{Chandra} data processing is the same as described earlier for A1795.       

\section{Discussion: Possible origins of the cavity}

One obvious possibility is that the cavity has just formed due to AGN feedback from the core of A1795 as its jets have interacted with the ICM, and that projection effects prevent us from seeing the cavity on the opposite side of the core. However this interpretation has some complications. The energy stored in the cavity is particularly high (it is $3.8^{2}_{-1}\times10^{60}$ erg, such that $PV\simeq10^{60}$ erg), and would be expected to produce a strong surface brightness contrast. All of the other clusters having cavities with similar or higher energies (see \citealt{Rafferty2006}, \citealt{Larrondo2012}) are observed to have pairs of cavities: MS 0735.6+7421 ($PV=8\times10^{60}$ erg), Zw 2701 ($PV=2\times10^{60}$ erg), Zw 3146 ($PV=2\times10^{60}$ erg), MACS J0913.7+4056 ($PV=1.2\times10^{60}$ erg). 

One possibility is that the X-ray bright, 46 kpc long filament extending to the south east is preventing us from seeing the inner edge of the companion bubble, which may have risen in this direction. As bubbles rise from the centre of the cluster, they can move along gravitational equipotentials relatively easily without any further energy input, so the location of the companion bubble does not necessarily have to be directly opposite to the north western bubble. For example, in the Perseus cluster (\citealt{Fabian2006}), the ghost bubbles to the north west and south are not aligned exactly on opposite sides of the cluster, but may be the result of the same outburst event due to their similar distances from the BCG. This may be the result of precession of the jets inflating the bubbles, while the two bubbles on either side of the cluster core detach at different times (\citealt{Dunn2006}). 

The lack of radio emission associated with the north western cavity (Fig. \ref{radiooverplot}) may mean that it is an old `ghost' cavity similar to those found in, for example, Perseus (\citealt{Fabian2006}). Such cavities are believed to be sufficiently old that the relativistic electron population can no longer radiate high frequency radio emission. 

A possible bubble to the south of the core, at the end of the H$\alpha$ and X-ray filament (shown by the white arrows in Fig. \ref{sectors}), was reported in \citet{Crawford2005}. However, as discussed earlier in section \ref{sec:cavitysize}, the cavity is abnormally small given its distance from the core (it has an unusually small radius of 3.5kpc), and should have expanded to a much greater volume if it has risen from the core. If the southern depression is a bubble which has risen from the core and dragged behind it the H$\alpha$ and X-ray filament, it would be difficult to explain why bubbles rising in other directions would not also drag H$\alpha$ and X-ray filaments behind them. The absence of such filaments between the core and the north western cavity is in tension with the idea that this cavity has originated from feedback from the central AGN. 

The metal abundance distribution (Fig. \ref{metals}), also shows there to be no metal enrichment coincident with the northwestern cavity, which is also in tension with the idea that it has risen from the cluster centre. In fact, the metals are displaced to the opposite side of the core to the northwestern cavity. One possibility is that the sloshing activity in the cluster, which has lead to the formation of the southern cold front reported in \citet{Markevitch2001}, has been the dominant process in the movement of metals, and has prevented the rising northwestern cavity from uplifting metals.  

If the northwestern cavity does have a roughly identical companion which cannot be seen due to the prominent southern filament, we can calculate whether the power associated with the outburst producing the cavity could have uplifted the gas in the filament, which has a total hot gas mass of $5 \times 10^{9}$ M$_{\odot}$ (\citealt{Fabian2001}). We can calculate a simple upper estimate of the gravitational energy of the uplifted gas in the filament, $U$, as
\begin{eqnarray}
   U= \frac{GM_{\rm cool}M(R)}{R}
   \end{eqnarray}   
where R is the length of the filament, $M(R)=10^{13}$M$_{\odot}$ is the total cluster mass interior to the filament, and $M_{\rm cool}$=$5 \times 10^{9}$ M$_{\odot}$ is the gas mass of the filament. The resulting energy is $U=8 \times 10^{58}$ erg. Using the age of the cavity calculated earlier (58 Myr), the power needed is $3.7\times10^{43}$ erg s$^{-1}$, which is nearly a factor of 60 smaller than the power needed to inflate the northwestern cavity calculated earlier ($2.1\times10^{45}$ erg s$^{-1}$). We therefore see that the power of this AGN outburst would easily have been able to uplift the gas of the southern filament. The energy needed to uplift the gas in the filament is a small fraction of the total enthalpy of the north western cavity. Therefore we would expect that the work done by any companion bubble in uplifting the filament would not significantly alter its dimensions compared to the north western cavity.

Given the suggestions that the northwestern cavity may not have risen from the core (i.e. its low abundance, and lack of filaments rising in its direction), one dramatic alternate way of forming the cavity could theoretically be through a past powerful outburst from the galaxy located inside it. Inflating the cavity in this way would naturally account for the low metal abundance, the lack of filaments extending to the cavity, and absence of radio emission. The total enthalpy of the northwestern cavity (the total of the energy contained within it and the work done against the surrounding ICM in inflating it), $4 \times 10^{60}$ erg, corresponds to the energy released by accretion onto a black hole of $2 \times 10^{7}$ M$_{\odot}$ of mass, assuming an efficiency of 10 percent for the conversion of rest mass into energy. This is significantly larger than the central black hole mass estimated in \citet{Maksym2014} for the galaxy WINGS J1348 inside the cavity, which is believed to lie in the range log$[M_{\rm BH}/{\rm M_{\odot}}] \sim 5.3- 5.7$. It is also nearly ten percent of the mass of the whole galaxy ($M_{\rm gal}=3\times10^{8}$M$_{\odot}$). 

It is possible that an outburst from the galaxy could have been `blown back' as it encountered the denser ICM nearer the core, which may have distorted the inner edge of the cavity feature we can see and increased its radius of curvature. An estimate of the lower limit to the cavity size can be obtained by using the volume enclosed by a sphere centred on the galaxy and reaching out to the inner rim of the cavity, shown earlier as the cyan circle in the top right hand panel in Fig. \ref{image}, which has a radius of 15.8 kpc. Using this volume, the lower limit to the total enthalpy of the cavity is $4 \times 10^{59}$ erg, corresponding to the energy released by accretion of $2 \times 10^{6}$ M$_{\odot}$ of mass onto a black hole. This is still significantly larger than the central black hole mass estimated in \citet{Maksym2014} for the galaxy inside the cavity. It is therefore very challenging to envisage a scenario in which an outburst from this galaxy could have led to the formation of such a large cavity. 

One possibility could be that the outburst from the galaxy has had such a dramatic effect on the galaxy that accurately determining the black hole mass using conventional means, such as the the black hole mass to bulge mass relationship (M$_{BH}$ $-$ M$_{bulge}$) or the black hole mass to velocity dispersion relationship (M$_{BH} -\sigma$), is not possible. It is also unclear how reliable the M$_{BH}$ $-$ M$_{bulge}$ and M$_{BH} -\sigma$ relations are for black hole masses below 10$^{6}$ M$_{\odot}$, as these relations are constrained mostly using results from higher mass black holes (\citealt{Kormendy2013}).

%\section{Conclusions}  

%\section{Summary}
%
%
%    
%
%\label{summary}

\section*{Acknowledgements}

SAW and ACF acknowledge support from ERC Advanced
Grant FEEDBACK. This
work is based on observations obtained with the \emph{Chandra} observatory, a NASA mission.
We thank Simona Giacintucci for providing the VLA radio contours presented in Fig. \ref{radiooverplot}.

\bibliographystyle{mn2e}
\bibliography{A1795_paper}

\appendix

\section[]{}

\label{sec:appendix}

\begin{table*}
  \begin{center}
  \caption{On axis ACIS-S \emph{Chandra} observations used. }
  \label{ACIS-S_obs}
  
    \leavevmode
    \begin{tabular}{lllllll} \hline \hline
    Obs ID & Detector&  Exposure / ks & RA & DEC & Date\\ \hline
494	&	ACIS-S	&	19.52	&	13 48 52.70	&	+26 35 27.00	&	1999-Dec-20  \\
493	&	ACIS-S	&	19.63	&	13 48 52.70	&	+26 35 27.00	&	2000-Mar-21 \\
3666	&	ACIS-S	&	14.42	&	13 48 52.70	&	+26 35 27.00	&	2002-Jun-10  \\
5286	&	ACIS-S	&	14.29	&	13 48 52.70	&	+26 35 27.00	&	2004-Jan-14  \\
5287	&	ACIS-S	&	14.3	&	13 48 52.70	&	+26 35 27.00	&	2004-Jan-14  \\
5288	&	ACIS-S	&	14.57	&	13 48 52.70	&	+26 35 27.00	&	2004-Jan-16  \\
6160	&	ACIS-S	&	14.84	&	13 48 52.70	&	+26 35 27.00	&	2005-Mar-20  \\
10900	&	ACIS-S	&	15.82	&	13 48 52.70	&	+26 35 27.00	&	2009-Apr-20  \\
10901	&	ACIS-S	&	15.47	&	13 48 52.70	&	+26 35 27.00	&	2009-Apr-20  \\
12028	&	ACIS-S	&	14.97	&	13 48 52.70	&	+26 35 27.00	&	2010-May-10  \\
12029	&	ACIS-S	&	14.69	&	13 48 52.70	&	+26 35 27.00	&	2010-Apr-28  \\
13106	&	ACIS-S	&	9.91	&	13 48 52.70	&	+26 35 27.00	&	2011-Apr-01  \\
13107	&	ACIS-S	&	9.64	&	13 48 52.70	&	+26 35 27.00	&	2011-Apr-01  \\
14268	&	ACIS-S	&	9.93	&	13 48 52.70	&	+26 35 27.00	&	2012-Mar-26  \\
14269	&	ACIS-S	&	9.94	&	13 48 52.70	&	+26 35 27.00	&	2012-Apr-08  \\
15485	&	ACIS-S	&	9.94	&	13 48 52.70	&	+26 35 27.00	&	2013-Apr-21  \\
15486	&	ACIS-S	&	9.68	&	13 48 52.70	&	+26 35 27.00	&	2013-Apr-22  \\
16432	&	ACIS-S	&	9.94	&	13 48 52.70	&	+26 35 27.00	&	2014-Apr-02  \\ \hline
  & Total exposure & 241.5 & & &  \\    \hline

    \end{tabular}
  \end{center}
\end{table*}

\begin{table*}
  \begin{center}
  \caption{On axis ACIS-I \emph{Chandra} observations used. }
  \label{ACIS-I_obs}
  \begin{tabular}{lllllll} \hline \hline
    Obs ID & Detector&  Exposure / ks & RA & DEC & Date\\ \hline
5289	&	ACIS-I	&	14.95	&	13 48 52.70	&	+26 35 27.00	&	2004-Jan-18  \\
5290	&	ACIS-I	&	14.95	&	13 48 52.70	&	+26 35 27.00	&	2004-Jan-23  \\
6159	&	ACIS-I	&	14.85	&	13 48 52.70	&	+26 35 27.00	&	2005-Mar-20  \\
6161	&	ACIS-I	&	13.59	&	13 48 52.70	&	+26 35 27.00	&	2005-Mar-28  \\
6162	&	ACIS-I	&	13.6	&	13 48 52.70	&	+26 35 27.00	&	2005-Mar-28  \\
6163	&	ACIS-I	&	14.85	&	13 48 52.70	&	+26 35 27.00	&	2005-Mar-31  \\
10898	&	ACIS-I	&	15.74	&	13 48 52.70	&	+26 35 27.00	&	2009-Apr-20  \\
10899	&	ACIS-I	&	14.92	&	13 48 52.70	&	+26 35 27.00	&	2009-Apr-22  \\
12026	&	ACIS-I	&	14.92	&	13 48 52.70	&	+26 35 27.00	&	2010-May-11  \\
12027	&	ACIS-I	&	14.85	&	13 48 52.70	&	+26 35 27.00	&	2010-Mar-16  \\
13108	&	ACIS-I	&	14.86	&	13 48 52.70	&	+26 35 27.00	&	2011-Mar-10  \\
13109	&	ACIS-I	&	14.58	&	13 48 52.70	&	+26 35 27.00	&	2011-Mar-11  \\
13110	&	ACIS-I	&	14.58	&	13 48 52.70	&	+26 35 27.00	&	2011-Mar-11  \\
13111	&	ACIS-I	&	14.58	&	13 48 52.70	&	+26 35 27.00	&	2011-Mar-11  \\
13112	&	ACIS-I	&	14.58	&	13 48 52.70	&	+26 35 27.00	&	2011-Mar-11  \\
13113	&	ACIS-I	&	14.58	&	13 48 52.70	&	+26 35 27.00	&	2011-Mar-11  \\
14270	&	ACIS-I	&	14.28	&	13 48 52.70	&	+26 35 27.00	&	2012-Mar-25  \\
14271	&	ACIS-I	&	13.98	&	13 48 52.70	&	+26 35 27.00	&	2012-Mar-25  \\
14272	&	ACIS-I	&	14.58	&	13 48 52.70	&	+26 35 27.00	&	2012-Mar-25  \\
14273	&	ACIS-I	&	14.58	&	13 48 52.70	&	+26 35 27.00	&	2012-Mar-26  \\
14274	&	ACIS-I	&	14.88	&	13 48 52.70	&	+26 35 27.00	&	2012-Apr-02  \\
14275	&	ACIS-I	&	14.88	&	13 48 52.70	&	+26 35 27.00	&	2012-Apr-07  \\
15487	&	ACIS-I	&	14.89	&	13 48 52.70	&	+26 35 27.00	&	2013-Jun-02  \\
15488	&	ACIS-I	&	14.58	&	13 48 52.70	&	+26 35 27.00	&	2013-Apr-10  \\
15489	&	ACIS-I	&	14.58	&	13 48 52.70	&	+26 35 27.00	&	2013-Apr-08  \\
15490	&	ACIS-I	&	14.89	&	13 48 52.70	&	+26 35 27.00	&	2013-Apr-17  \\
15491	&	ACIS-I	&	14.89	&	13 48 52.70	&	+26 35 27.00	&	2013-Apr-18  \\
15492	&	ACIS-I	&	14.58	&	13 48 52.70	&	+26 35 27.00	&	2013-Apr-15  \\
16434	&	ACIS-I	&	14.89	&	13 48 52.70	&	+26 35 27.00	&	2014-Apr-02  \\
16435	&	ACIS-I	&	14.58	&	13 48 52.70	&	+26 35 27.00	&	2014-Apr-03  \\
16436	&	ACIS-I	&	14.58	&	13 48 52.70	&	+26 35 27.00	&	2014-Apr-03 \\
16437	&	ACIS-I	&	14.59	&	13 48 52.70	&	+26 35 27.00	&	2014-Apr-03  \\    \hline
  & Total exposure & 469.2 & & &  \\    \hline

    \end{tabular}
  \end{center}
\end{table*}

%
%\appendix
%\section[]{}
%\label{sec:appendix}
%
%
%
%
%
%
%\clearpage

%\section[]{ROSAT imaging in E11}

\end{document}